\documentclass{PoS}

\usepackage{listings}
\usepackage{color}
\usepackage{amsmath}
\definecolor{gray}{rgb}{0.4,0.4,0.4}
\definecolor{darkblue}{rgb}{0.0,0.0,0.6}
\definecolor{cyan}{rgb}{0.0,0.6,0.6}
\def\bea{\begin{eqnarray}}
\def\eea{\end{eqnarray}}

\title{Trace anomaly and dynamical quark mass}

\ShortTitle{Dynamical quark mass}

\author{\speaker{Yi-Bo Yang}\thanks{We thank the RBC and UKQCD collaborations for providing
us their DWF gauge configurations. The computer resources of this work is supported by the Strategic Priority Research Program of Chinese Academy of Sciences (Grant No. XDC01040100). P. Sun is supported by Natural Science Foundation of China under grant No. 11975127, {Z. Liu is supported by Natural Science Foundation of China under grant No. 11935017}, Y. Yang is supported in part by Chinese Academy of Science CAS Pioneer Hundred Talents Program.}\\
        CAS Key Laboratory of Theoretical Physics, Institute of Theoretical Physics, Chinese Academy of Sciences, Beijing 100190, China\\
        }
        
\author{Jian Liang\\
        Department of Physics and Astronomy, University of Kentucky, Lexington, KY 40506, USA\\}

\author{Zhaofeng Liu\\
        Institute of High Energy Physics, Chinese Academy of Sciences, Beijing 100049, China\\}
      
\author{Peng Sun\\
        Nanjing Normal University, Nanjing, Jiangsu, 210023, China\\}

\abstract{
\begin{center}
\large{
\vspace*{0.4cm}
\includegraphics[scale=0.20]{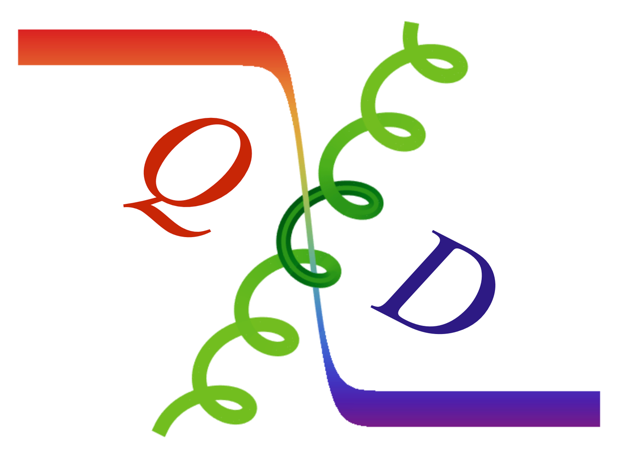}\\
\vspace*{0.4cm}
($\chi$QCD Collaboration)
}
\end{center}
We investigated the origin of the RI'/MOM quark mass under the Landau gauge at the non-perturbative scale, using the chiral fermion with different quark masses and lattice spacings. Our result confirms that such a mass is non-vanishing based on the linear extrapolation to the chiral and continuum limit, and shows that such a mass comes from the spontaneous chiral symmetry breaking induced by the near zero modes with the eigenvalue $\lambda<{\cal O}(5m_q)$, and is proportional to the quark matrix element of the trace anomaly at least down to $\sim $1.3 GeV.}

\FullConference{37th International Symposium on Lattice Field Theory - Lattice2019\\
		16-22 June 2019\\
		Wuhan, China}

\begin{document}

\section{Background}

The QCD energy momentum tensor (EMT)  in the classical level
\bea
T_{\mu\nu}=T_{q,\mu\nu}+T_{{g},\mu\nu},\ \ T_{{q},\mu\nu}=\frac{1}{4}\bar{\psi}\gamma_{(\mu}\overleftrightarrow{D}_{\nu)}\psi,\ \  T_{g,\mu\nu}=F_{\mu\rho}F^{\ \ \rho}_{\nu}-\frac{1}{4}g_{\mu\nu}F^2,
\eea
which is traceless up to the quark mass term $m\bar{\psi}\psi$, can have a non-trivial trace after the UV regularization is applied. As demonstrated in Ref. \cite{Hatta:2018ina}, $\frac{1}{4}$ but not $\frac{1}{4-2\epsilon}$ in front of $F^2$ leaves a residual trace $-\frac{\epsilon}{2}F^2$ under the dimensional regularization, and then the total trace of EMT with the quantum correction of $F^2$ becomes
\bea\label{eq:anomaly}
T^{\alpha}_{\ \alpha}=m\bar{\psi}\psi+T^{\alpha}_{a,\alpha},\ \  T^{\alpha}_{a,\alpha}=\gamma_mm\bar{\psi}\psi+\frac{\beta}{2g}F^2.
\eea
where $T^{\alpha}_{a,\alpha}$ is the trace anomaly term. 

Since the EMT in a hadron state should satisfy the relation $\langle P|T_{\mu\nu}|P\rangle=\frac{P_{\mu}P_{\nu}-\frac{1}{4}P^2}{P_0}$ where $|P\rangle$ is the hadron state with 4-momentum $P$, one will have the following equation in the rest frame of the hadron state $|H\rangle$ with mass $M_H$,
\bea\label{eq:mass}
M_H=\langle H|T^{\alpha}_{\ \alpha}|H\rangle=\langle H|m\bar{\psi}\psi|H\rangle+\langle H|T^{\alpha}_{a,\alpha}|H\rangle.
\eea
With the practical calculation of $\langle H|m\bar{\psi}\psi|H\rangle$ based on lattice QCD or phenomenology inputs, one can predict that $T^{\alpha}_{a,\alpha}$ contributes most of the nucleon mass~\cite{Shifman:1978zn}. One of the goals of the proposed Electron-Ion Collider (EIC) is to test this prediction with the photoproduction $\gamma N \rightarrow J/\psi N$ near the threshold, which would be sensitive to the form factor of $F^2$~\cite{Hatta:2018ina}. 

On the other hand, one can have another understanding on the nucleon mass based on the quark model. In the quark model, the nucleon mass majorly comes from the sum of the ``dynamical" mass of three component quarks which is $\sim$ 300 MeV each. At a perturbative scale like 2 GeV, the $u/d$ averaged current quark mass can be accurately determined by the lattice QCD calculation as $\sim$3.4 MeV in  $\overline{\textrm{MS}}$ scheme (see FLAG2019~\cite{Aoki:2019cca}): The bare quark mass $m$ corresponding to the physical pion mass 139 MeV, multiplies the non-perturbative renormalization constant $Z_m$ in the RI'/MOM scheme at scale $Q$ and also a fixed-order perturbative matching between RI'/MOM(Q) and $\overline{\textrm{MS}}$(2 GeV). The dependence of the intermedia scale $Q$ will be cancelled up to the discretization error ${\cal O}(a^2Q^2)$, when $Q$ is in the perturbative region. But the RI'/MOM calculation of $Z_m m$ also shows that there would be a ``residual quark mass" which is $\sim300$ MeV in the limit $m_q\rightarrow0$ and $Q\rightarrow0$~\cite{Bowman:2005vx}, even though one can not convert it to the $\overline{\textrm{MS}}$ scheme since the convergence of perturbative matching is poor at small $Q$. Such a residual quark mass is considered as a character of the spontaneous chiral symmetry breaking ($\chi$SB), and also a hint of the dynamical quark mass in the phenomenological quark model~\cite{Bhagwat:2007vx}. 

A {natural} guess to connect the two pictures above is that the RI'/MOM residual quark mass is related to the trace anomaly in the quark state. It is somehow counterintuitive since such a matrix element under the dimensional regularization should be proportional to the bare quark mass based on the perturbative calculation as the first term of $T^{\alpha}_{a,\alpha}$, and then vanish in the chiral limit. But the perturbative calculation also shows that the $Z_mm$ under the dimensional regularization should vanish in the chiral limit, unlike that in the RI'/MOM scheme under the Landau gauge. In this proceeding, we will show that the spontaneous $\chi$SB induced by the near zero modes of the chiral fermion, makes both the residual RI'/MOM quark mass and the trace anomaly in the quark state to be non-zero, around the chiral limit with small $Q$.

\section{Simulation setup}

In this calculation, we use the overlap valence fermions on the $(2+1)$ flavor RBC/UKQCD DWF+Iwasaki configurations using four ensembles at $a=$0.11 fm with pion masses 139, 330, 400 and 530 MeV respectively, and also one ensemble at $a=$0.084 fm with physical pion mass 139 MeV to control the discretization error, as shown in Table~\ref{table:lat}. In the study of the spontaneous $\chi$SB, using the chiral fermion satisfying $\{D_c, \gamma_5\} = 0$ \cite{Chiu:1998gp} through the overlap fermion approach is essential to avoid any additional chiral symmetry breaking in most of the lattice fermion actions, and the detailed implementation of the overlap fermion can be found in previous work of $\chi$QCD collaboration~\cite{Li:2010pw}.

\begin{table}[htbp]
\begin{center}
\caption{\label{table:lat} The parameters for the RBC/UKQCD configurations~\cite{Blum:2014tka}: spatial/temporal size, lattice spacing,  strange and light quark mass under $\overline{\textrm{MS}}$ scheme at {2 GeV}, pion mass with the degenerate light sea quark, and the number of configurations.}
\begin{tabular}{llccccr}
Symbol & $L^3\times T$  &a (fm)  &$m_s$(MeV)&$m_l$(MeV)&  {$m_{\pi}$}(MeV)   & $N_{cfg} $ \\
\hline
64I &$64^3\times 128$& 0.0837(2) & 95  & 4 &139 & 40  \\
48I &$48^3\times 96$& 0.1141(2) & 95  & 4 &139 & 40  \\
24I  & $24^3\times 64$& 0.1105(3) &120  &25  &330  & 203  \\
24Ih  & $24^3\times 64$& 0.1105(3) &120  &40 &400  & 143  \\
24Ih2 & $24^3\times 64$& 0.1105(3) &120  &71 &530  & 85  \\
\hline
\end{tabular}
\end{center}
\end{table}

\begin{figure}
  \centering
  \includegraphics[width=0.45\textwidth]{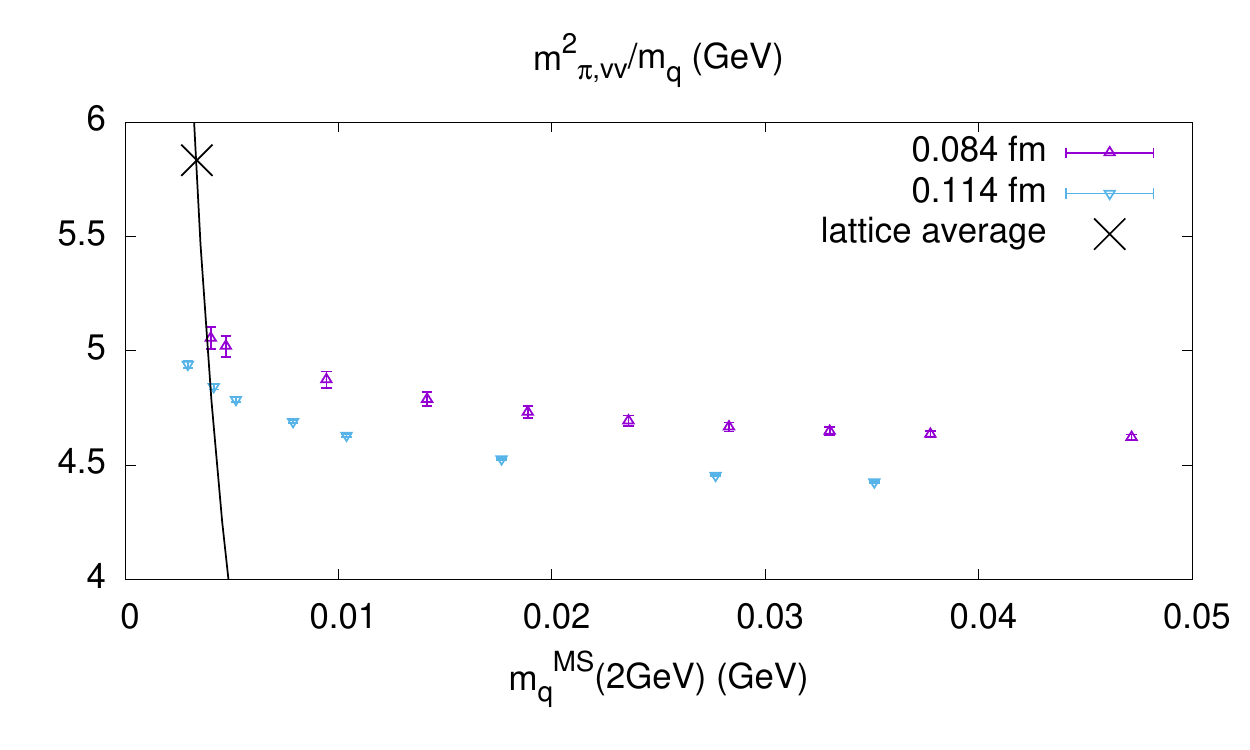}
   \includegraphics[width=0.45\textwidth]{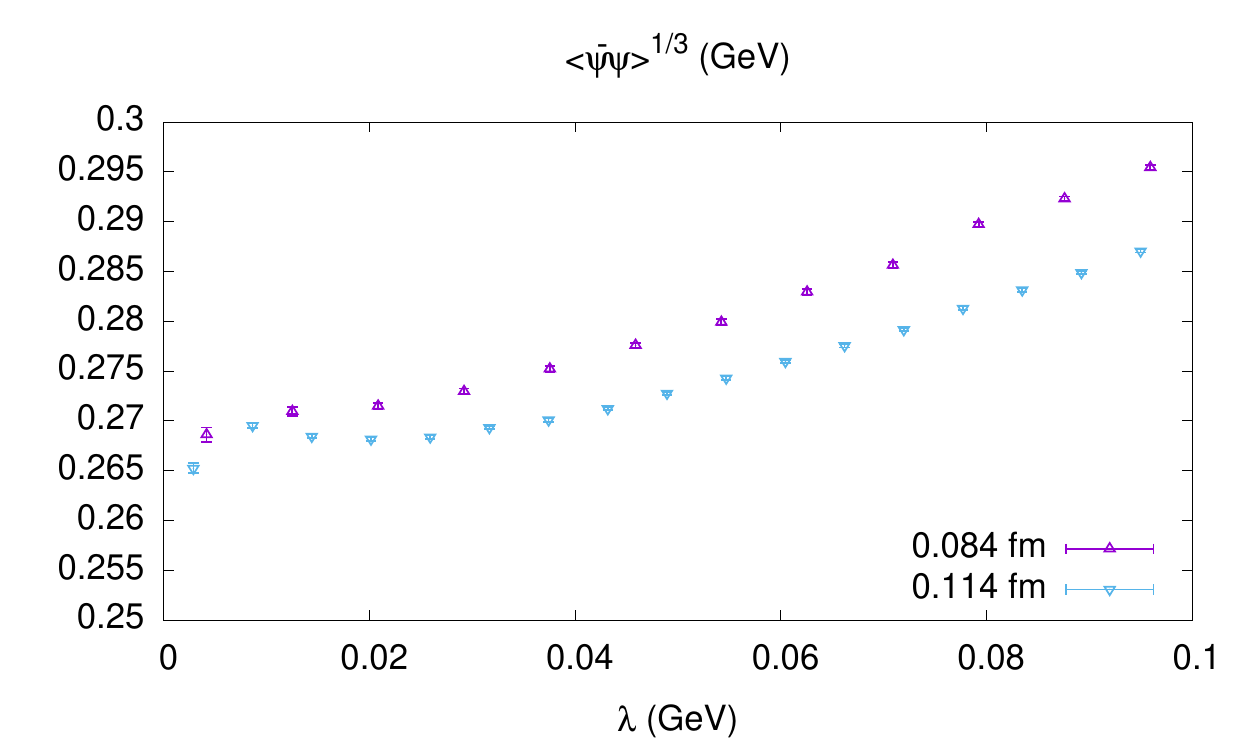}
  \caption{The partially quenched ratio $m_{\pi, vv}^2/m_q$ (left panel), and the eigenvalue density $(\frac{1}{\pi}\frac{\textrm{d} n(\lambda)}{\textrm{d} |\lambda|})^{1/3}\simeq \langle \bar{\psi}\psi\rangle^{1/3}$ with $\lambda$ being the eigenvalue of $D_c$ (right panel), at $\overline{\textrm{MS}}$ 2 GeV. The data from 48I (a=0.114fm) and that from 64I (a=0.084 fm) are the green and purple triangles respectively. The black line in the left panel shows the curve $(m^{phys}_{\pi})^2/m_q$ with $m^{phys}_{\pi}$=139 MeV, and its cross with the 48I/64I data points correspond to the physical quark mass on that ensemble.}
  \label{fig:test}
\end{figure}

Unlike the ensembles at 0.11fm, the 64I ensemble at 0.084fm has not been used for the overlap fermion calculation before this proceeding. So we would like to mention a few details here. With one step of HYP smearing, we use 800 H-Wilson eigenvectors under the eigenvalue upper band 0.158 and 1600 $D_c$ eigenvector pairs in the range $[-0.0795 i, 0.0795i]$ for efficient deflations in both the construction and inversion of the overlap fermion. The production cost on the 64I ensemble increases by roughly a factor of 5 compared with that on 48I. In Fig.~\ref{fig:test}, {we shown the partially quenched ratio $m_{\pi, vv}^2/m_q$ in the left panel, where the cross of the black line and the 48I/64I data points (in blue/purple) corresponds to the physical quark mass on that ensemble; and the the $D_c$ eigenvalue density $(\frac{1}{\pi}\frac{\textrm{d} n(\lambda)}{\textrm{d} |\lambda|})^{1/3}$ is shown in the right panel, it corresponds to the $\langle \bar{\psi}\psi\rangle^{1/3}$ at the $\lambda\rightarrow 0$ limit up to the volume correction and the other systematic uncertainties}. If we just do a trivial linear $a^2$ extrapolation, the $u/d$ averaged physical quark mass $m_q$ will be roughly 3.8 MeV, and the chiral condensate in the vacuum in the chiral limit would be $\langle\bar{\psi}\psi\rangle\sim$(270 MeV)$^3$. The details of the chiral condensate calculation will be presented in a separated work.

\section{Methodology}

The RI'/MOM renormalization requires the calculation of both the off-shell quark propagator and also the amputated vertex function $\Lambda(p, \Gamma)$. $\Lambda(p, \Gamma)$ in an off-shell quark state $|q\rangle$ under the Landau gauge condition can be implemented by
\bea \label{eq:wallsource}
\Lambda(p, \Gamma)=\frac{1}{V}S(p)^{-1}\cdot \left\langle \sum_w {\gamma_5}S ^{\dagger}(p,w){\gamma_5}\Gamma S(p,w) \right\rangle\cdot S(p)^{-1} \ ,
\eea
where $S(p)$ is volume source propagator
\bea
	S(p)=\frac{1}{V}\sum_{x} e^{-ipx}S(p,x),\ S(p,x)= \langle  \psi(x) \sum_{y}{\bar{\psi}(y)e^{ipy}}\rangle ,
\eea
and the summation over all lattice sites $w$ or $x$ are normalized by the 4-D volume $V$. 

\begin{figure}
  \centering
  \includegraphics[width=0.8\textwidth]{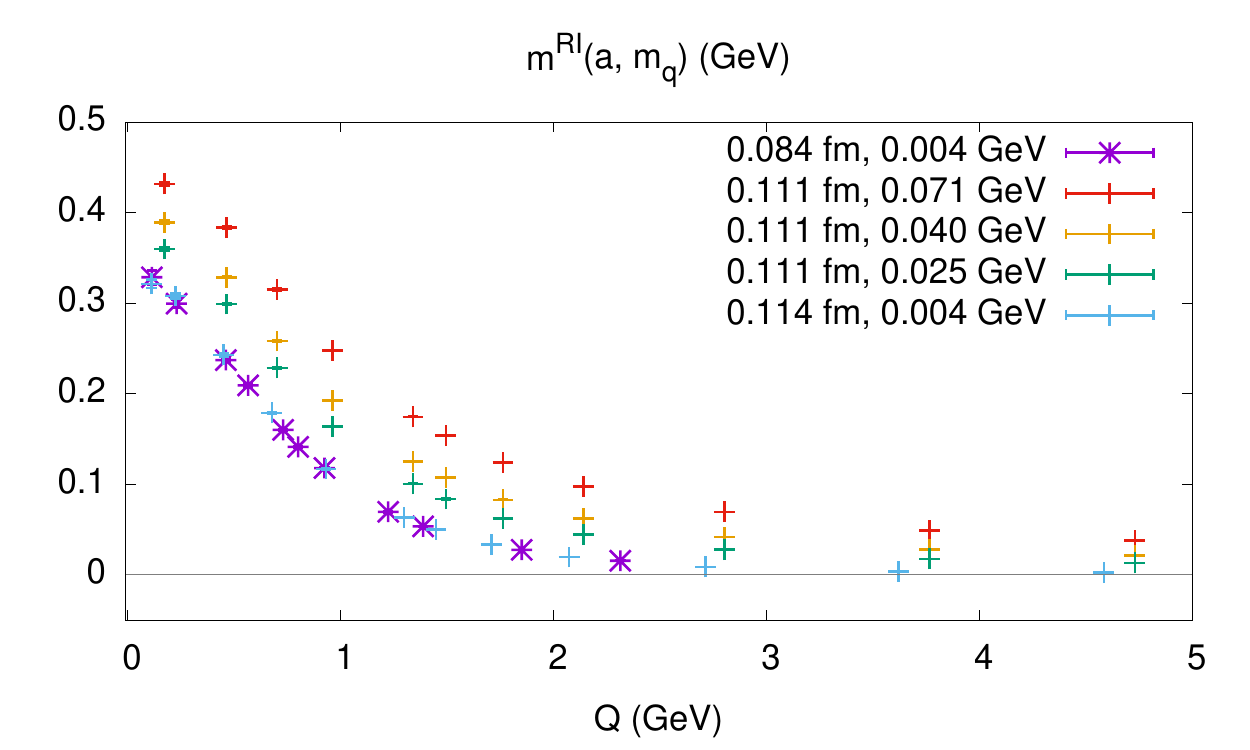}
  \caption{The RI'/MOM quark mass with different bare quark mass, as a function of scale $Q$. It is obvious that there is a residual mass after a linear extrapolation to the chiral limit, especially at small $Q$. The purple stars and blue crosses corresponds to the data at two different lattice spacings, and their consistency shows that the discretization error is small.}
  \label{fig:mass_scale}
\end{figure}

Eventually the {scalar, pseudoscalar, quark mass and quark field} renormalization constants at the RI'/MOM scale $Q$ are defined by~\cite{Bi:2017ybi}
\bea
&&\bar{Z}_S(Q,m_q)\equiv \frac{12Z_q(Q)}{\textrm{Tr}[\Lambda(p, I)]_{p^2=Q^2}}=\frac{A_S}{m^2_q}+Z_S(Q)+B_S*m_q+{\cal O}(m_q^2), \\
&& \bar{Z}_P(Q,m_q)\equiv \frac{12Z_q(Q)}{\textrm{Tr}[\gamma_5\Lambda(p, \gamma_5]_{p^2=Q^2}}=(\frac{A_P}{m_q}+Z^{-1}_P(Q)+B_P*m_q+{\cal O}(m_q^2))^{-1}, \\
&& Z_m(Q)=\frac{1}{12Z_q(Q)m_q}\textrm{Tr}[S(p)^{-1}]_{p^2=Q^2},\ Z_q(Q)=\frac{1}{12p^2}\textrm{Tr}[\sum_{\mu} \gamma_{\mu}p_{\mu}S(p)^{-1}]_{p^2=Q^2},
\eea
where $m_q$ is the quark mass, $Z_S=Z_P$ for the chiral fermion after the quark mass dependence is subtracted, and $A_S$ is small as shown in Ref.~\cite{Bi:2017ybi}. It is remarkable that the relation $Z_m\bar{Z}_P=1$ is exactly satisfied for all the momenta and quark masses based on the above definition, and immediately induces that
\bea
m^{RI}\equiv Z_m(Q)m_q=A_P+Z^{-1}_S(Q)m_q+{\cal O}(m_q^2),
\eea
where $A_P$ corresponds to the residual RI'/MOM mass which is independent on $m_q$, as illustrated in Fig.~\ref{fig:mass_scale}. 

Similarly, we can calculate the $F^2$ matrix element in the quark state,
\bea
\langle q|F^2|q\rangle(Q)= \frac{\textrm{Tr}[\frac{1}{V}S(p)^{-1}\cdot \langle F^2 S(p) \rangle\cdot S(p)^{-1}]_{p^2=Q^2}}{12Z_q(Q)},
\eea
and check whether it vanishes in the chiral limit. This matrix element is noisy due to the disconnected insertion of the gluon operator, thus we shall apply the CDER technique introduced in Ref.~\cite{Liu:2017man,Yang:2018nqn} to improve the signal. 

To investigate the role of the near zero modes of $D_c$ in this breaking, we can investigate the role of the near zero modes of $D_c$ by replacing the propagator $S(p,x)$ with its high mode part,
\bea
S^{h}(p,x,\lambda_c)=\sum_{y} e^{ipy}(\langle \bar{\psi}(x) \psi(y)\rangle-\sum_{-\lambda_c<\lambda<\lambda_c} \textrm{v}_{\lambda}(x)\frac{1}{i\lambda+m}\textrm{v}_{\lambda}^{\dagger}(y)),
\eea 
where $i\lambda$ and $\textrm{v}_{\lambda}$ are the eigenvalue and eigenvector of $D_c$ satisfying $D_c \textrm{v}_{\lambda}=i\lambda \textrm{v}_{\lambda}$, and consider all the above RI'/MOM calculation as a function of $\lambda_c$. The similar attempt has been made for the $\chi$SB in the meson correlators~\cite{Denissenya:2014ywa}, and it was found that the scalar and pseudo scalar meson correlators become degenerate after certain $\lambda_c$ is applied. In this proceeding, we will also check the quark mass dependence of $\lambda_c$ which can eliminate the spontaneous $\chi$SB.

\section{Preliminary result and summary}

\begin{figure}
  \centering
  \includegraphics[width=0.45\textwidth]{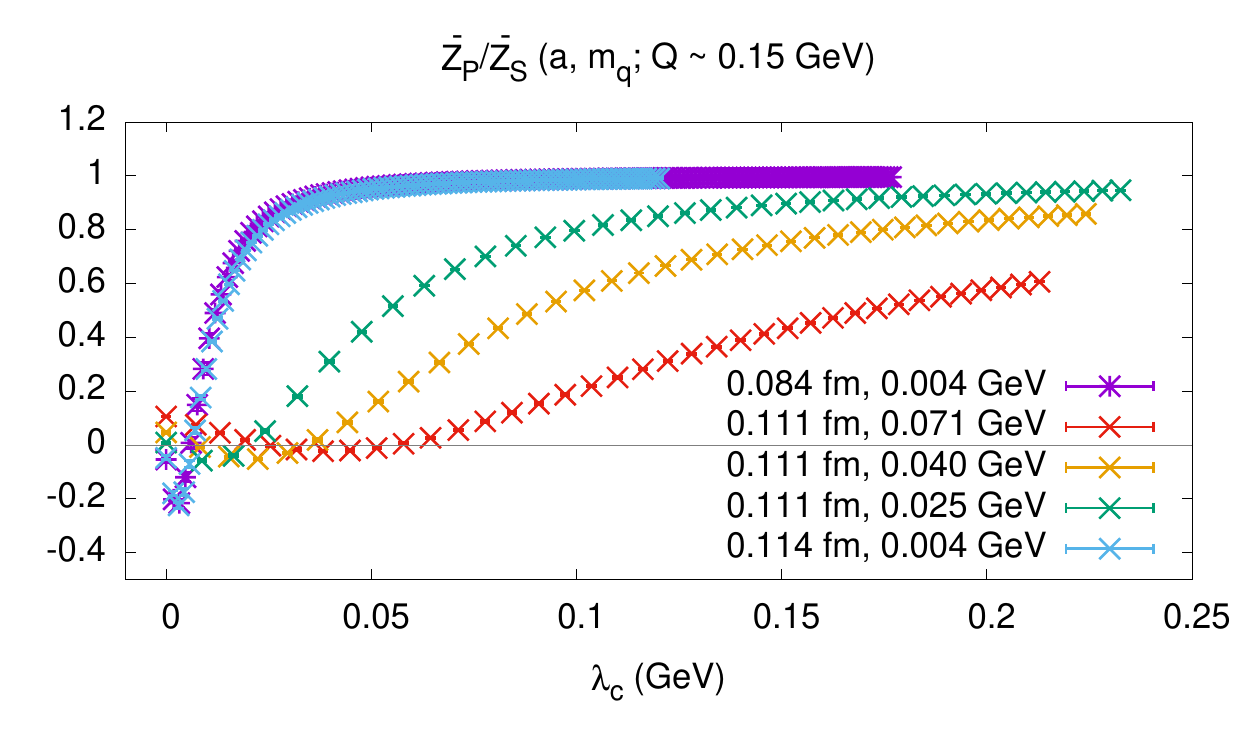}
   \includegraphics[width=0.45\textwidth]{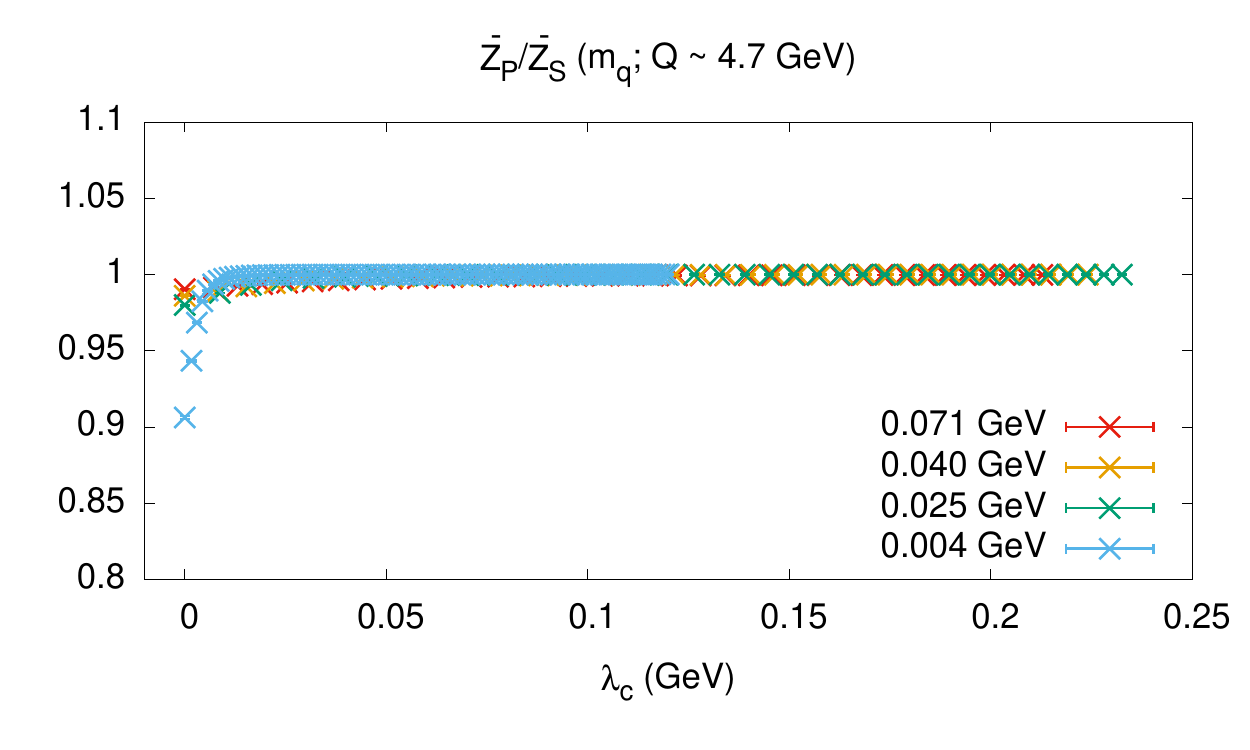}
  \caption{The $\chi$SB as a function of the eigenvalue cutoff $\lambda_c$. {Different colors are used for the data with different lattice spacings and quark masses}.}
  \label{fig:chiral}
\end{figure}

The eigenvalue cutoff $\lambda_c$ dependence of $\bar{Z}_P/\bar{Z}_S$ at two scales, 0.15 GeV and 4.7 GeV, are shown in Fig.~\ref{fig:chiral}. The case with $\lambda_c=0$ corresponds to the case using the full propagator with all the low modes. It is obvious that the breaking will appear when the low modes with eigenvalues smaller than ${\cal O}(5m_q)$ are included and the scale $Q\sim 0$, while the effect becomes much smaller when $Q$ is higher. 

\begin{figure}
  \centering
   \includegraphics[width=0.45\textwidth]{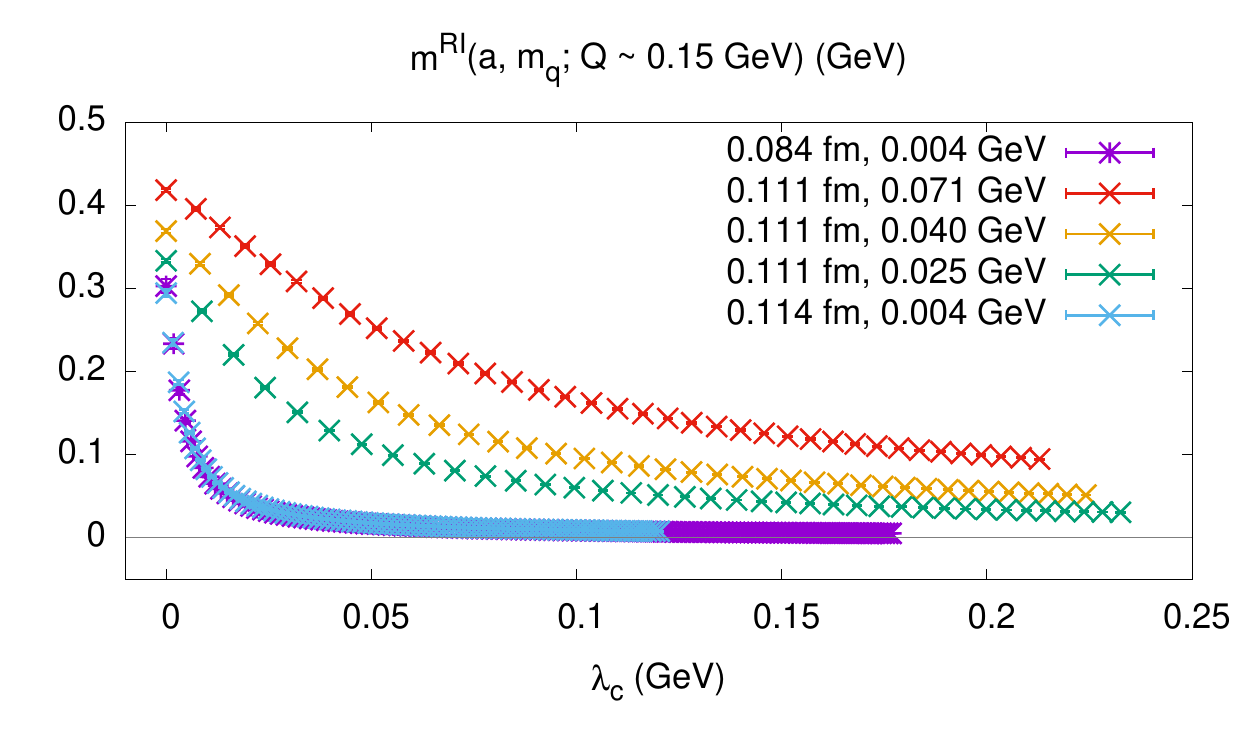}
   \includegraphics[width=0.45\textwidth]{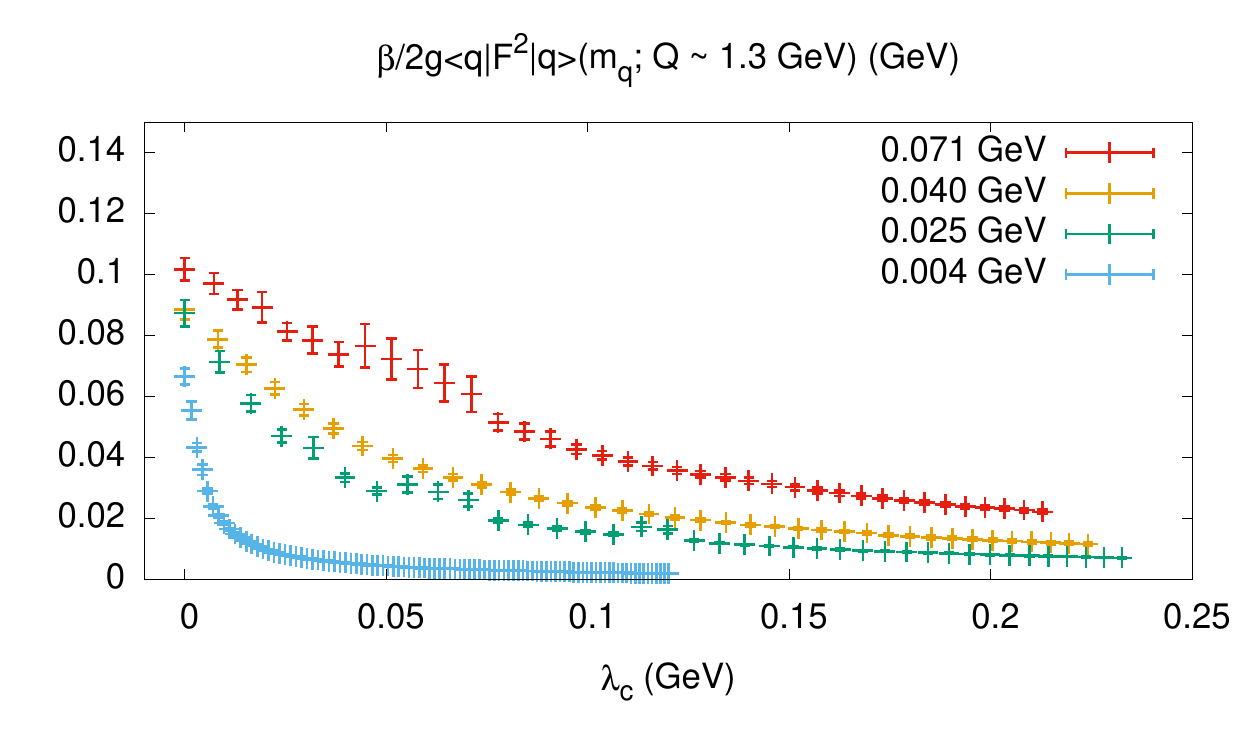}
  \caption{The RI'/MOM quark mass and matrix element $\frac{\beta}{2g}\langle q|F^2|q\rangle$ as the function of eigenvalue cutoff $\lambda_c$. }
  \label{fig:mass}
\end{figure}

Fig.~\ref{fig:mass} shows the $\lambda_c$ dependence of $m^{RI}$ and the gluon matrix element $\langle q|F^2|q\rangle$ with a factor $\frac{\beta}{2g}\sim0.03$. Similarly, only the near zero modes have the significant contribution to these two quantities, and the efficient cutoff to eliminate the spontaneous $\chi$SB is proportional to the quark mass. Note that a CDER cutoff at $\sim$ 1.5 fm is applied in the gluon case and the right panel of Fig.~\ref{fig:mass} shows the result at a higher scale $Q\sim1.3$ GeV, since less aggressive CDER cutoff should be applied with lower $Q$ and then the uncertainty is still large there with present statistics. 
 
\begin{figure}
  \centering
   \includegraphics[width=0.8\textwidth]{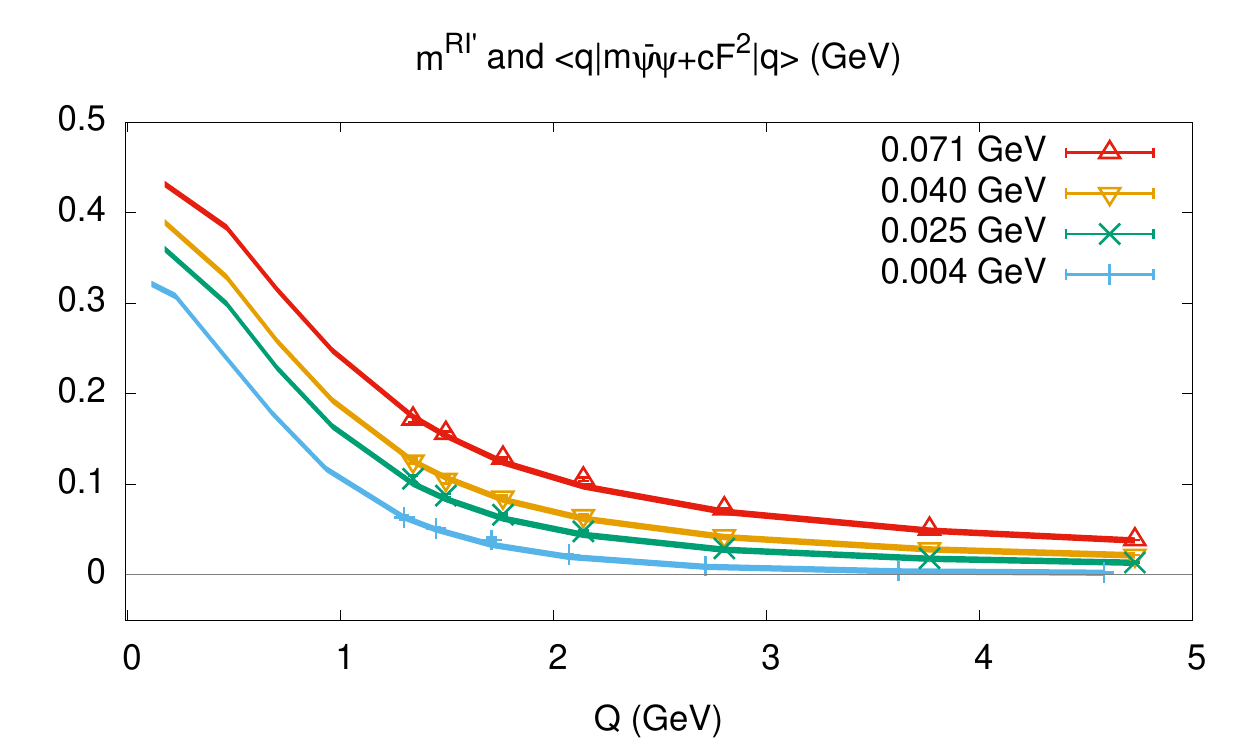}
  \caption{The combination $\langle q|m\bar{\psi}\psi+cF^2|q\rangle$ (colored data points) comparing with the RI'/MOM quark mass $m^{RI}$ (interpolated curves), as the function of energy scale $Q$. Different colors are used for the cases with different bare quark masses.}
  \label{fig:q2}
\end{figure}

We finally show the preliminary comparison of the RI'/MOM quark mass $m^{RI}$ and the combination $\langle q|m\bar{\psi}\psi+cF^2|q\rangle$ in Fig.~\ref{fig:q2}. With $c=0.023$ which is close to $\frac{\beta}{2g}$, their difference down to $Q\sim 1.3$ GeV is not observable. Thus the spontaneous $\chi$SB induced by the near zero modes actually makes both $m^{RI}$ and $\langle q|F^2|q\rangle$ to be non-zero around the chiral limit. Further studies would uncover the relation between two essential characters of QCD: spontaneous $\chi$SB and confinement. 

\bibliographystyle{unsrt}
\bibliography{trace_anomaly.bib}

\end{document}